\documentclass[lettersize,journal]{IEEEtran}
\usepackage{amsmath,amsfonts,bm,xcolor}
\usepackage{balance}
\usepackage{algorithmic}
\usepackage{resizegather}
\usepackage{algorithm}
\usepackage{array}
\usepackage[caption=false,font=normalsize,labelfont=sf,textfont=sf]{subfig}
\usepackage{textcomp}
\usepackage{csquotes}
\usepackage{stfloats}
\usepackage{tabstackengine}
\usepackage{url}
\usepackage{verbatim}
\usepackage{graphicx}
\usepackage{cite}
\usepackage{amssymb,amsmath,amsthm,enumitem}

\usepackage[cmintegrals]{newtxmath}
\begin{document}

\title{Channel Prediction Using Deep Recurrent Neural Network with EVT-based Adaptive Quantile Loss Function}

\author{Niloofar~Mehrnia,~\IEEEmembership{}
        Parmida~Valiahdi,~\IEEEmembership{}
        Sinem~Coleri,~\IEEEmembership{}% <-this % stops a space
        James~Gross~\IEEEmembership{}
\thanks{Niloofar Mehrnia and James Gross are with the Department of Electrical Engineering and Computer Sciences, KTH Royal Institute of Technology, Stockholm, Sweden (e-mail:\{nilome,jamesgr\}@kth.se). (Corresponding author: Niloofar Mehrnia)}
\thanks{Parmida Valiahdi and Sinem Coleri are with the Department of Electrical and Electronics Engineering,
Koc University, Istanbul, Turkey (e-mail:\{pvaliahdi23,scoleri\}@ku.edu.tr).}
\thanks{Sinem Coleri acknowledges the support of the Scientific and Technological Research Council of Turkey $2247$-A National Leaders Research Grant $\#121C314$.}
% <-this % stops a space
}

%\markboth{Journal of \LaTeX\ Class Files,~Vol.~, No.~, January~2025}%{Shell \MakeLowercase{\textit{et al.}}: A Sample Article Using IEEEtran.cls for IEEE Journals}

%\IEEEpubid{}

\maketitle

\begin{abstract}
Ultra-reliable low latency communication (URLLC) systems are pivotal for applications demanding high reliability and low latency, such as autonomous vehicles. In such contexts, channel prediction becomes essential to maintaining communication quality, as it allows the system to anticipate and mitigate the effects of fast-fading channels, thereby reducing the risk of packet loss and latency spikes. This letter presents a novel framework that integrates neural networks with extreme value theory (EVT) to enhance channel prediction, focusing on predicting extreme channel events that challenge URLLC performance. We propose an EVT-based adaptive quantile loss function that integrates EVT into the loss function of the deep recurrent neural networks (DRNNs) with gated recurrent units (GRUs) to predict extreme channel conditions efficiently. 
The numerical results indicate that the proposed GRU model, utilizing the EVT-based adaptive quantile loss function, significantly outperforms the traditional GRU. It predicts a tail portion of $7.26\%$, which closely aligns with the empirical $7.49\%$, while the traditional GRU model only predicts $2.4\%$. This demonstrates the superior capability of the proposed model in capturing tail values that are critical for URLLC systems.
\end{abstract}

\begin{IEEEkeywords}
Channel prediction, deep recurrent neural network, extreme value theory, URLLC.
\end{IEEEkeywords}

\section{Introduction}
\label{sec:introduction}
\IEEEPARstart{A}{chieving} ultra-reliable low latency communication (URLLC) in mission-critical
applications necessitates effective real-time channel prediction, especially when dealing with extreme events, i.e., very low values happening infrequently, to mitigate the effects of fast-fading channels. Accurate channel predictions enable dynamic adjustments to modulation and coding scheme (MCS) selection, optimizing throughput and minimizing delays associated with channel quality indicators (CQI) for proactive tasks. This improves resource allocation efficiency in $5$G and beyond \cite{urllc_mcs_01}.
URLLC ensures the delivery of critical information with a packet error rate ranging from $10^{-9}$ to $10^{-5}$, and delays on the order of milliseconds. 
Neural networks are promising for achieving high prediction accuracy through the processing of extensive real-time data, thereby enhancing URLLC performance. To refine these predictions further, integrating extreme value theory (EVT) with neural network techniques can effectively analyze rare fading events that affect ultra-reliable communications. By concentrating on the lower tail of signal power distributions, EVT offers precise modeling of wireless channels under extreme conditions. Therefore, the combination of EVT and neural networks represents a robust strategy for improving channel prediction in URLLC systems.

Several advanced neural network architectures have been proposed to improve channel prediction accuracy \cite{channelprediction_nn_04, channelprediction_nn_01, channelprediction_nn_02}. Recurrent neural networks (RNNs) leverage temporal correlations in sequential data for channel prediction \cite{channelprediction_nn_02}, unlike convolutional neural networks (CNNs) that use matrices as input, thereby ignoring temporal correlations \cite{channelprediction_nn_04}. However, RNNs face challenges like vanishing or exploding gradients, which have been mitigated by using long short-term memory (LSTM) or gated recurrent units (GRU) \cite{channelprediction_nn_01, channelprediction_nn_02}. In \cite{channelprediction_nn_urllc_01}, deep RNNs (DRNNs) combined with LSTM or GRU have been proposed to predict CQI under URLLC latency constraints, improving reliability without reducing data rates. Despite these advancements, most studies focus on average channel prediction and fail to address extreme events critical for URLLC. 

EVT provides a robust framework for modeling extreme events by leveraging the generalized Pareto distribution (GPD) to estimate tail statistics, crucial for ultra-reliability \cite{MehrniaTWC, MehrniaTWC2,bennisAoI}. EVT has been applied to wireless channel modeling \cite{MehrniaTWC}, including extensions like multivariate EVT (MEVT) for spatial diversity \cite{MehrniaTWC2}, and queue-aware resource allocation \cite{bennisAoI}. However, these studies focus on tail estimation using maximum likelihood estimation (MLE) and assume large data availability. Recently, \cite{channelestimation_nn_evt_urllc_01} proposed a generative adversarial network (GAN)-based GPD parameter estimator to accurately estimate GPD parameters. GAN has consistently demonstrated superior performance compared to MLE in estimating the probability density function (PDF) of extreme events, especially with limited sample sizes. Nonetheless, none of these works consider real-time channel prediction of both average statistics and extreme values, crucial for URLLC performance.

In this letter, we propose a novel channel prediction framework incorporating EVT into DRNN to address the reliability and latency constraints of URLLC. Unlike previous DRNN applications focusing on frequent average statistic predictions, our approach leverages EVT-based adaptive quantile loss function to predict extreme values over the next few milliseconds. Specifically, we use DRNN with GRU, which is computationally more efficient than DRNN with LSTM for channel prediction and holds promise for future real-world implementations \cite{channelprediction_nn_urllc_01}. To show the superiority of the proposed scheme, we use real channel measurements to ensure extreme values are detected over the next few milliseconds.

\section{System Model}
\label{sec:system_model}
\subsection{System Model Assumptions}
\label{sec:system_model_assumps}
We examine a real-time URLLC system designed to predict current and future received power samples based on historical data, particularly under conditions of severe fading that result in extremely low received power levels. In this context, the transmitter (Tx) transmits packets to a receiver (Rx) at a constant power over an unknown channel. As a result, the received signal power can be interpreted as the squared amplitude of the channel state information.
Initially, the receiver focuses on estimating the channel, emphasizing extreme values. This estimation process can utilize either dedicated pilot signals and training sequences or the historical data from previous transmissions along with corresponding feedback. The Rx gathers channel measurements and subsequently estimates the channel tail distribution. This estimated distribution serves as prior knowledge, which is then relayed back to the Tx to facilitate channel prediction using a DRNN with GRU architecture. 
%\vspace{-0.4cm}
\subsection{Channel Prediction}
\label{sec:system_model_pred}
The communication channel between Tx and Rx is characterized as a time-varying fading channel. Let $\bm{x}_t$ represent the transmitted signal at the $t$-th time instant, and denote the channel fade coefficient at the same instant by $\bm{h}_t$. The received signal at the receiver is denoted as $\bm{y}_t$, while $\bm{n}_t$ represents the additive noise. Consequently, the relationship can be expressed as follows $\bm{y}_t = \bm{h}_t \, \bm{x}_t + \bm{n}_t$.
The channel coefficients $\bm{h}_t$ are statistically modeled using the pilot signals or previous data transmissions. Feedback and processing delays can render the channel coefficients obtained at the Tx outdated by the time they are utilized, leading to significant performance degradation in adaptive transmission systems. Channel prediction aims to determine a predicted value, $\hat{\bm{h}}_{t+\tau}$, at time $t$ in Tx side that approximates its actual value at the upcoming time $t+\tau$ as close as possible, i.e., $\hat{\bm{h}}_{t+\tau} \approx \bm{h}_{t+\tau}$.

The time-varying channel coefficients exhibit temporal correlation. By modeling this correlation, the number of pilot signals required for estimating these coefficients can be reduced. The learned model can then be applied for channel prediction. However, since the correlation model may vary across different channel geometries, a purely statistical approach may lack robustness. To address this issue, we propose a deep learning-based solution to learn the channel correlation model and predict future channel coefficients, thereby minimizing the need for pilot transmissions. Specifically, we utilize a DRNN with GRU architecture to develop the proposed channel predictor. In the subsequent subsection, we provide a brief overview of DRNN, with a particular focus on the GRU component.
%\vspace{-0.4cm}
\subsection{Deep Recurrent Neural Network for Channel Prediction}
\label{sec:system_model_drnn}
\begin{figure}
    \centering    \includegraphics[width=8cm,height=5cm]{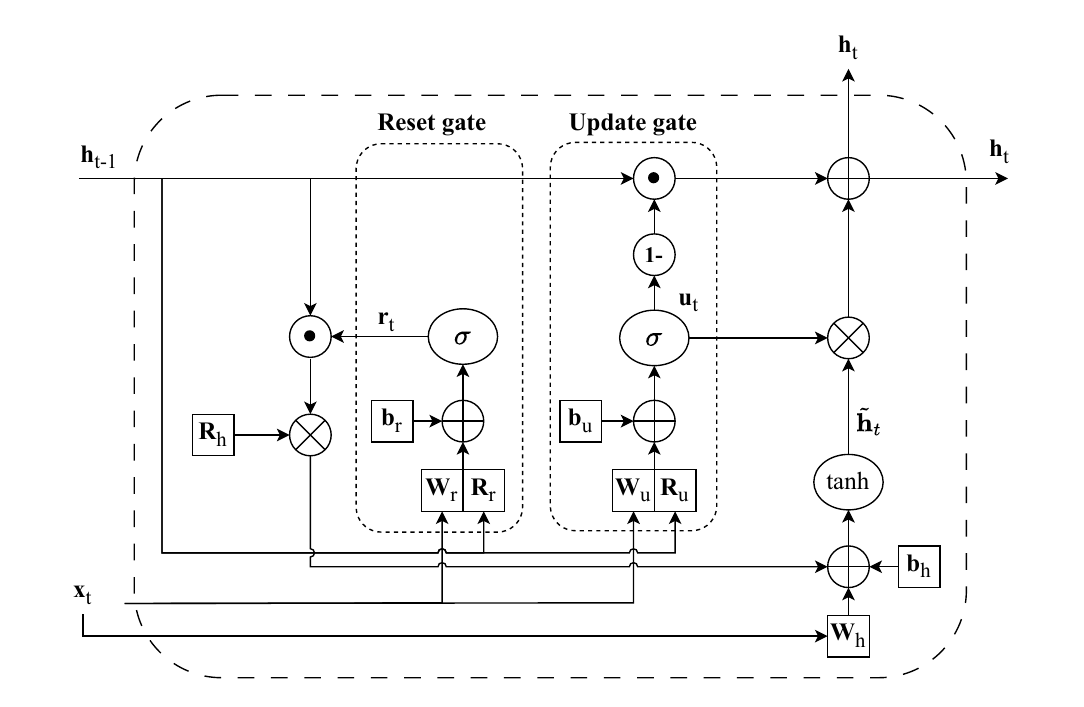}
    %\vspace{-3mm}
    \caption{The architecture of GRU.}
    \label{fig:gruarch}
\end{figure}

In the training phase, the system model uses DRNN with GRU to effectively capture temporal dependencies in time series data. Then, in the operation phase, it uses the pre-trained model to predict the current and future received power values, especially the extreme ones. At each time step $t \in \{1,\dots,T\}$, the output of GRU, $\hat{\bm{y}}_t$, is computed as 
\begin{equation}
\label{eqn:gruoutput}
    \hat{\bm{y}}_t = g(\bm{W}_{y} \bm{h}_t + \bm{b}_y), 
\end{equation}
where $\bm{b}_y$ is the bias vector in the output layer and $g$ is the output activation function and is given by 
\begin{equation}
    g(\bm{k}) = \text{max} \{ \Re[\bm{k}],0 \} + j \, \text{max} \{ \Im [\bm{k}],o \},
\end{equation}
with $\Re[.]$ and $\Im[.]$ being the real and imaginary parts of the vectors, respectively. In (\ref{eqn:gruoutput}), the hidden units, $\bm{h}_t$, is connected to the output layer, $\hat{\bm{y}}_t$, with the weighted connections $\bm{W}_{y}$. 
The GRU architecture consists of a total of two gates: a reset gate that combines the current input with previous memory and an update gate, that determines the percentage of previous state to remember, accompanied by a single hidden state. Within the GRU, each gate is equipped with a sigmoid activation function, while a separate \enquote{tanh} function is employed to generate the output, as depicted in Fig.~\ref{fig:gruarch}. 
The hidden state $\bm{h}_t$ in (\ref{eqn:gruoutput}) gets updated based on the current input $\bm{x}_t$ and the previous hidden state $\bm{h}_{t-1}$ as follows:
%\vspace{-3mm}
\begin{subequations}\label{eqn:grunn}
\begin{align}
\bm{r}_t &= \sigma_g (\bm{W}_{r} \bm{x}_t + \bm{R}_{r} \bm{h}_{t-1} + \bm{b}_r), \label{eqn:grunn2}\\
\bm{u}_t &= \sigma_g (\bm{W}_{u} \bm{x}_t + \bm{R}_{u} \bm{h}_{t-1} + \bm{b}_u), \label{eqn:grunn1}\\
\tilde{\bm{h}}_t &= \text{tanh} (\bm{W}_{h} \bm{x}_t + \bm{R}_{h} (\bm{r}_t \odot \bm{h}_{t-1}) + \bm{b}_h), \label{eqn:grunn3}\\
\bm{h}_t &= (\bm{1}-\bm{u}_t) \odot \bm{h}_{t-1} + \bm{u}_t  \tilde{\bm{h}}_t, \label{eqn:grunn4}
\end{align}
\end{subequations}
where $\bm{r}_t$ and $\bm{u}_t$ represent the reset gate and update gate for a time stamp $t$, respectively; $\bm{W}_r$ and $\bm{W}_u$ are the weight matrices for the input state at the reset gate and update gate, respectively; $\bm{R}_r$ and $\bm{R}_u$ represent the weight matrices for the hidden state at the reset gate and update gate, respectively; $\bm{b}_r$, $\bm{b}_u$, and $\bm{b}_h$ refer to the bias terms at the reset gate, update gate, and hidden state, respectively; and $\odot$ denotes Hadamard product of the matrix. $\bm{h}_{t-1}$ represents the previous hidden state of the GRU, and $\sigma_g$ is the sigmoid activation function given by $\sigma_g (\bm{x}) = 1/(1+e^{-\bm{x}})$.
To quantify the difference between the predicted values $\hat{\bm{y}}_t$ and the actual values $\bm{y}_t$, we need to define a loss function $\mathcal{L} (\bm{y}_t,\hat{\bm{y}}_t)$. By employing a structured approach to weight updates based on feedback from the loss function, the model iteratively minimizes the loss such that 
\begin{align}
W^* &= \underset {\bm{W},\bm{R},\bm{b}}
{\text{arg min}} \hspace{5pt} \mathcal{L}(\bm{y},\hat{\bm{y}}).
\end{align}
Here, $\mathcal{L}(\bm{y},\hat{\bm{y}}) = \sum_{t=1}^{T} \mathcal{L}(\bm{y}_t,\hat{\bm{y}}_t)$ is an overall summation of losses in each timestep $\mathcal{L}(\bm{y}_t,\hat{\bm{y}}_t)$, and is differentiable to find the optimal parameters $\bm{W} \in \{ \bm{W}_r, \bm{W}_u \}$, $\bm{R} \in \{ \bm{R}_r, \bm{R}_u \}$, and $\bm{b} \in \{ \bm{b}_r, \bm{b}_u, \bm{b}_h \}$, leading to improved predictions. An iterative procedure persists until the minimum achievable error between the estimated value $\hat{\bm{y}}$ and the actual value $\bm{y}$ is attained.
Common loss functions like mean-squared error (MSE) and mean absolute error (MAE) are widely focused on average error metrics prediction. 
Although DRNNs, including GRU, have demonstrated strong capabilities in predicting the next received power of a wireless channel, they often struggle with predicting extreme values due to their infrequent occurrence.
 
%\vspace{-0.3cm}
\section{EVT-based Adaptive Quantile Loss Function}
\label{sec:methodology}
In this section, we define an EVT-based adaptive quantile loss function using the estimated GPD parameters fitted to the channel tail distribution. This loss function provides better feedback to the DRNN with the GRU model for predicting extreme events.

%\vspace{-4mm}
\subsection{EVT-based Channel Estimation}
\label{sec:methodology_estimation}
In the channel tail estimation stage, the receiver applies EVT-based channel estimation following \cite[Theorem~$1$]{MehrniaTWC} and uses the received power values to estimate the channel tail distribution. The channel measurements are converted into a sequence of $n$ independent and identically distributed (i.i.d.) samples with cumulative distribution function (CDF) $F$ by removing dependencies using declustering \cite{MehrniaTWC}. This sequence is denoted by $X^n = \{x_1,\dots , x_n\}$, where $x_i$ denotes the $i^{th}$ sample, $i \in \{1,...,n\}$. The GPD is then fitted to the received power values exceeding a threshold to model the channel tail distribution as:
\begin{equation}
\label{eqn:gpddist}
\small
    F(z) = 1-[1+\frac{\xi z}{\tilde{\sigma}_{u}}]^{-1/\xi},
    \normalsize
\end{equation}
where $z$ is a non-negative value denoting the exceedance below threshold $u$, i.e., $z=u-X$ for $X=\{x_i<u\}$; $\xi$ and $\tilde{\sigma}_{u}=\sigma+\xi(u-\mu)$ are the shape and scale parameters of the GPD, respectively; and $\mu$ and $\sigma$ are the location and scale parameters of the generalized extreme value (GEV) distribution fitted to the CDF of $M_n = \max \{ x_1,...,x_n\}$ \cite[Theorem~1]{MehrniaTWC}, \cite{evt_04}. The optimal threshold $u$ is determined using two complementary methods: mean residual life (MRL) and parameter stability methods \cite{MehrniaTWC, evt_04}. This optimum threshold distinguishes extreme values from non-extreme ones and depends on the data distribution. The estimation of Pareto distribution parameters is performed using the GAN-EVT method introduced in \cite{channelestimation_nn_evt_urllc_01}, which has been shown to outperform the GPD parameters estimation using MLE.
We assume the distribution of received powers is stationary, meaning the parameters defining the GPD distribution remain constant. If non-stationarity is detected using the Augmented Dickey-Fuller (ADF) test, external factors influencing the GPD parameters are identified. Based on these factors, the dataset is then divided into smaller, stationary groups for further analysis. The GPD parameters are then estimated for the sequence over which the channel can be considered stationary, as explained in \cite{MehrniaTVT}.
%\vspace{-3mm}
\subsection{GRU with Adaptive Quantile Loss Function for Channel Prediction}\label{sec:methodology_prediction}
In our EVT-based channel prediction framework, initially, the channel tail distribution is estimated using GPD. The estimated parameters are incorporated into the loss function of the GRU. This allows the model to adaptively adjust the desired quantile value, improving its ability to predict extreme values over the next few milliseconds.

The quantile loss function is a promising candidate for integrating the prediction of the extreme quantiles into neural networks \cite{qloss_02}. 
The quantile loss function $\mathcal{L}_{q} (\bm{y}_t,\hat{\bm{y}}_t) = q  (\bm{y}_t - \hat{\bm{y}}_t)$ if $\bm{y}_t > \hat{\bm{y}}_t$ and otherwise, $\mathcal{L}_{q} (\bm{y}_t,\hat{\bm{y}}_t) =(1 - q)  (\hat{\bm{y}}_t - \bm{y}_t)$, for a given quantile $q$ \cite{qloss_02}.
The loss is asymmetric, penalizing overestimates and underestimates differently depending on $q$. Traditional quantile regression targets a specific quantile, which can lead to an overestimation of extreme quantiles \cite{qloss_02}. In contrast, the proposed EVT-based adaptive quantile loss function adjusts the $q$ value using the GPD parameters in (\ref{eqn:gpddist}) fitted to the channel tail distribution, enabling predictions for all values in the tail rather than a single quantile.

Let GPD($\sigma$,$\xi$) be the Pareto model fitted to the training samples $X^{n} = \{x_1,x_2,...,x_n\},$ $x_i<u$ for $i \in \{1,\dots,n\} $, where $\sigma$ and $\xi$ are the scale and shape parameters of the GPD model fitted to the tail samples, respectively. Then, the EVT-based adaptive loss function of the GRU is defined as:
\begin{equation}
  \mathcal{L}_{\alpha} (\bm{y}_t,\hat{\bm{y}}_t) = \begin{cases} \alpha  (\bm{y}_t - \hat{\bm{y}}_t) & \text{if }  \bm{y}_t > \hat{\bm{y}}_t, \hspace{5pt}  \bm{y}_t<u \\ (1 - \alpha)  (\hat{\bm{y}}_t - \bm{y}_t) & \text{if }  \bm{y}_t \leq \hat{\bm{y}}_t, \hspace{5pt} \bm{y}_t<u \end{cases}
   \label{eqn:lossfunc}
\end{equation}
where $\alpha = F(u-\bm{y}_t)  \Pr(\bm{y}_t<u)$. Here, $F(.)$ is the CDF of the GPD fitted to the channel tail distribution in (\ref{eqn:gpddist}), $u$ is the optimum threshold separating extreme values from the near-average values, and $\Pr(\bm{y}_t<u)$ denotes the probability of observing $\bm{y}_t$ less than $u$. In the case of $\bm{y}_t>u$, the loss function is the traditional MSE-based loss.
%Then, we show that $\hat{\bm{y}}_t = F^{-1}(\alpha)$ is the best prediction for $\bm{y}_t$ that results in the minimum of the loss function.
The objective is to find a prediction $\hat{\bm{y}}_t$ that minimizes the expected adaptive quantile loss, $\min\mathbb{E}[\mathcal{L}_{\alpha}(\bm{y}_t, \hat{\bm{y}}_t)]$, where $\alpha = F(u-\bm{y}_t)  \Pr(\bm{y}_t<u)$ for $\bm{y}_t<u$. The expected loss is given by
%\vspace{-2mm}
\begin{equation}\label{eqn:convexdf1}
\begin{aligned}
    \mathbb{E}[\mathcal{L}_{\alpha}(\bm{y}_t, \hat{\bm{y}}_t)] = 
    \int_{-\infty}^{\hat{\bm{y}}_t} (1-{\alpha})(\hat{\bm{y}}_t - \bm{y}_t) f(\bm{y}_t) \, d\bm{y}_t + \\
    \int_{\hat{\bm{y}}_t}^{\infty} {\alpha}(\bm{y}_t - \hat{\bm{y}}_t) f(\bm{y}_t) \, d\bm{y}_t,
\end{aligned}
\end{equation}
where $\mathbb{E}[.]$ denotes the expectation operator and $f(\bm{y}_t)$ is the probability density function of $\bm{y}_t$. To proceed further, we need to show that (\ref{eqn:convexdf1}) is convex. Therefore, we differentiate the function with respect to $\hat{\bm{y}}_t$ as follows:
\begin{equation}\label{eqn:convexdf2}
    \begin{aligned}
    \small
        \frac{d}{d\hat{\bm{y}}_t}  \mathbb{E}[\mathcal{L}_{\alpha}(\bm{y}_t, \hat{\bm{y}}_t)]  =
        (1-\alpha) \int_{-\infty}^{\hat{\bm{y}}_t} f(\bm{y}_t)\, d\bm{y}_t - \alpha \int_{\hat{\bm{y}}_t}^{\infty} f(\bm{y}_t) \, d\bm{y}_t.
    \end{aligned}
    \normalsize
\end{equation}
Taking the second derivative of (\ref{eqn:convexdf1}), we obtain
\begin{equation}\label{eqn:convexdf3}
    \begin{aligned}
        \frac{d^2}{d\hat{\bm{y}}^2_t}  \mathbb{E}[\mathcal{L}_{\alpha}(\bm{y}_t, \hat{\bm{y}}_t)]  =
        (1-\alpha)  f(\hat{\bm{y}}_t) + \alpha f(\hat{\bm{y}}_t) = f(\hat{\bm{y}}_t).
    \end{aligned}
\end{equation}
Since the results of (\ref{eqn:convexdf3}) is always non-negative, $\mathbb{E}[\mathcal{L}_{\alpha}(\bm{y}_t, \hat{\bm{y}}_t)]$ is convex and we can take its derivative with respect to \( \hat{\bm{y}}_t \) and set it to zero,
    $(1-{\alpha})F(\hat{\bm{y}}_t) - {\alpha}(1-F(\hat{\bm{y}}_t)) = 0$, where $F(\hat{\bm{y}}_t) = \int_{-\infty}^{\hat{\bm{y}}_t} f(\bm{y}_t) \, d\bm{y}_t$ is the CDF of $\bm{y}_t$. Since \( {\alpha} = F(u-\bm{y}_t)  \Pr(\bm{y}_t<u) \), simplifying this equation results in:
\begin{equation}
    F(\hat{\bm{y}}_t) = F(u-\bm{y}_t) \Pr(\bm{y}_t < u).
\end{equation}
Therefore, minimizing the EVT-based adaptive quantile loss function results in a prediction \( \hat{\bm{y}}_t \) that corresponds to the \( F(u-\bm{y}_t) \Pr(\bm{y}_t < u) \)-th quantile of the distribution. 
The intuition is that, for the tail values, if the prediction $\hat{\bm{y}}_t$ underestimates the observed value $\bm{y}_t$, the penalty should be proportional to $F(u-\bm{y}_t) \Pr(\bm{y}_t < u)$. Otherwise, if $\hat{\bm{y}}$ overestimates $\bm{y}_t$, the penalty should be proportional to $1- F(u-\bm{y}_t) \Pr(\bm{y}_t < u)$. Please note that the proposed EVT-based adaptive loss function assigns the same weight to the calculated loss for $\bm{y}_t>u$ and the loss calculated for $\bm{y}_t<u$.
%\vspace{-2mm}
\section{Numerical Results}
\label{sec:numerical_results}
This section aims to evaluate the effectiveness of the proposed methodology in predicting received power values for a system operating under URLLC conditions. We focus on assessing the ability of the proposed scheme to predict the extreme received powers in the next few milliseconds (msec). We also compare these results with those obtained using the traditional GRU model for channel prediction. To the best of our knowledge, DRNN models, especially GRU, with the traditional MSE loss function are the optimum learning approach for predicting the time series \cite{channelprediction_nn_urllc_01}. 
%\textcolor{blue}{It is worth noting that although the quantile regression neural network is suggested for predicting a specific quantile, since this model does not consider the temporal dependency among the samples, adding complexity, and predicting only a specific quantile, we skip to compare our proposed work with the quantile regression neural networks. That is why we have chosen only the GRU with MSE loss function as baselines for comparing our proposed approach.}  
%\vspace{-0.3cm}
\subsection{Data Characterization}
\label{sec:numerical_results_datachar}
The channel measurements were collected from the engine compartment of a Fiat Linea operating at $60$~GHz, a millimeter-wave frequency that holds significant potential for both automotive and indoor applications. The position of the transmitter and receiver antennas are selected to capture the effects of engine vibration on the received power. A vector network analyzer (VNA) (R\&S\textsuperscript{\textregistered} ZVA$67$) was connected to the antennas via R\&S\textsuperscript{\textregistered} ZV-Z$196$ port cables, which have a maximum transmission loss of $4.8$ dB. The horn antennas used for transmission and reception feature a nominal gain of $24$ dBi and a $12^\circ$ vertical beamwidth. The transmitter and receiver antenna locations were chosen from the available options for wireless sensors within the engine compartment, specifically at positions $5$ and $13$ as shown in \cite{vehicular_02}-Fig. 1. Approximately 
$10^6$ consecutive samples were recorded with a time resolution of $2$~msec. This resolution is sufficiently short to capture rapid fluctuations in the received signal strength caused by the engine vibrations. The proposed algorithm was implemented using Python.
%\vspace{-3mm}
\subsection{Model Characterization}
\label{sec:numerical_results_gruchar}
The proposed method employs a single GRU layer with $50$ units to capture temporal dependencies in the input data. We conducted ablation studies to explore different GRU configurations and determined that a single GRU layer with $50$ units achieved the best balance between performance and computational efficiency. Since the primary objective of this study is to compare the performance of GRU models using different loss functions, we selected this simple and efficient architecture. This choice ensures fast predictions while allowing us to focus on evaluating the impact of the loss functions on extreme value prediction. Compiled with the Adam optimizer at a learning rate of $0.0008$, it utilizes an EVT-based adaptive quantile loss function to enhance extreme value predictions. The model is trained for $20$ epochs with a batch size of $256$, using a validation set to ensure generalization for predicting the next $\{1, 2, 3, 4, 5\}$ samples, corresponding to a maximum prediction horizon of $10$~ms. Hyperparameters are tuned experimentally for optimal performance.
To prepare the dataset, we convert raw time-series data into a supervised format using a sliding window technique, extracting sequences of past observations. We selected look-back (LB) periods of $5$ or $15$ data points, with $5$ addressing the strict $10$~ms latency requirement of URLLC, while $15$ allows for greater prediction accuracy. The GRU with EVT-based adaptive quantile loss accurately predicts tail samples by utilizing GPD parameters estimated from excess values below a threshold of $-49$~dBm, derived from the GAN-EVT model \cite{channelestimation_nn_evt_urllc_01}. During prediction, the loss function for the values above the threshold is the MSE loss, while those below are predicted with the EVT-based adaptive quantile loss, guided by pre-assigned parameters.

\subsection{Discussion}
\label{sec:numerical_results_discussion}
\begin{figure}[h]
    \centering
    \includegraphics[width=7cm,height=4cm]{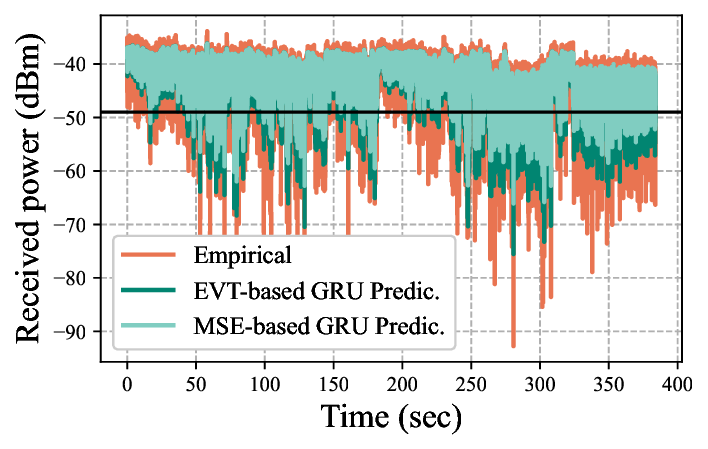}
    %\vspace{-4mm}
    \caption{GRU prediction using MSE-based and EVT-based adaptive quantile losses. The horizontal black line shows the optimum threshold of $-49$~dBm.}
    \label{fig:power_time}
\end{figure}

Fig.~\ref{fig:power_time} depicts the power values for the validation dataset, representing $20\%$ of the experimental data and using the last $5$ samples (LB $= 5$) for prediction. The proposed GRU model with EVT-based adaptive quantile loss outperforms the GRU model trained with MSE loss in predicting extreme values, particularly below the $-49$~dBm threshold. Specifically, the EVT-based model correctly predicts $21\%$ of values below $-49$~dBm, compared to $7\%$ for the MSE-based model. Both models have identical computational complexity, each containing $7950$ trainable parameters and requiring approximately $30$~$\mu$sec per prediction. Thus, the proposed adaptive quantile loss function does not increase model complexity.
%Fig.~\ref{fig:power_time} depicts the power values for the validation dataset, comprising $20\%$ of the entire experimental dataset and utilizing the last $5$ samples (LB $= 5$) for the current value prediction. The proposed model utilizing EVT-based adaptive quantile loss surpasses the performance of a GRU model trained with MSE loss in the prediction of extreme values. Fig.~\ref{fig:power_time} highlights the ability of the proposed model to predict samples below the $-49$ dBm threshold. The true predicted values below the threshold of $-49$~dBm for the GRU using the MSE loss function and the GRU with the EVT-based adaptive quantile loss function are $7\%$ and $21\%$, respectively, demonstrating the superior performance of the proposed loss function scheme. From the perspective of computational complexity, the time taken per prediction for the GRU model utilizing the MSE loss function is $30$~$\mu$sec, while the GRU model employing the proposed EVT-based adaptive quantile loss function requires $29$~$\mu$sec. These times are nearly identical, and both models share the same GRU and therefore, contain the same number of trainable parameters, totaling $7950$. Therefore, the implementation of the proposed adaptive quantile loss function does not introduce additional complexity to the model. 
When the LB is increased to $15$, the prediction delay grows by less than a factor of $2$, contrary to the expected linear scaling. This deviation can be attributed to TensorFlow's ability to parallelize certain operations, which mitigates the scaling effects. Additionally, fixed overhead costs, such as model loading, remain constant regardless of input length, further contributing to the sub-linear increase in delay.
%\vspace{-0.4cm}

\begin{figure}[h]
\centering
\captionsetup[subfigure]{labelformat=empty}
     \begin{center}
        \subfloat[(a)]{%
            \label{fig:hist_gruevt}
            \includegraphics[width=4.1cm,height=3cm]{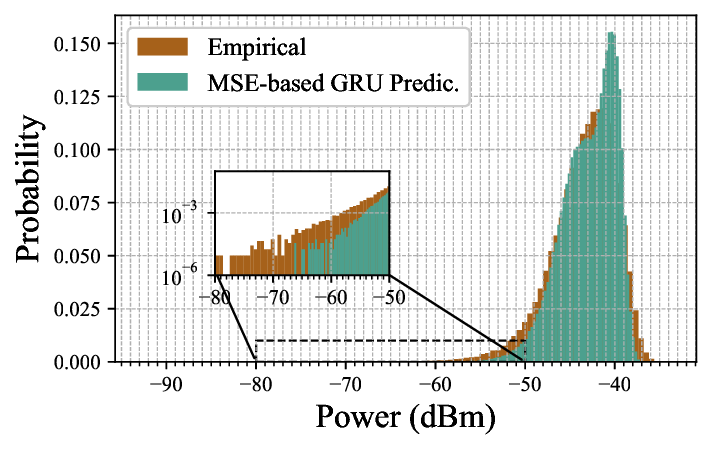}
        }\ 
        \subfloat[(b)]{%
            \label{fig:hist_gruevt}
            \includegraphics[width=4.1cm,height=3cm]{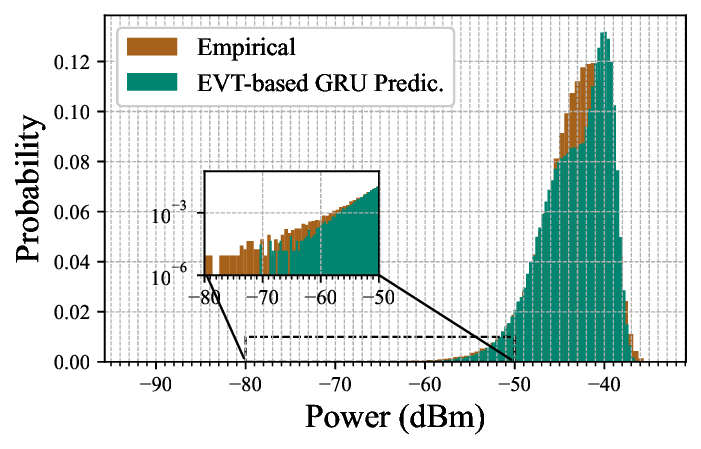}
        }\
    \end{center}
    %\vspace{-2mm}
    \caption{PDF comparison between the empirical values and GRU predicted values: (a) GRU with the MSE-based loss function, and (b) GRU with the EVT-based adaptive quantile loss function.}
   \label{fig:histogram}
\end{figure}

%Furthermore, the superior performance of the proposed scheme can be enhanced further by dynamically estimating the GPD parameters within the loss function and also by assigning different weights to the losses calculated for tail predictions and near-average value predictions, which are both directions for future work.}

Fig.~\ref{fig:histogram}  compares the PDFs of the empirical data with those predicted by the GRU using both EVT-based adaptive quantile loss and MSE loss functions. Both models effectively estimate average values. However, the GRU with the EVT-based adaptive quantile loss function outperforms in predicting tail distributions. It also provides more accurate predictions near the average values, with better scaling around the peak of the distribution. This is particularly evident in the zoomed-in sections of the tail, where the proposed model successfully captures the tail values that significantly influence the performance of URLLC systems.
The results show that the GRU with MSE loss predicts the tail portion of the distribution as $2.4\%$. In comparison, the GRU with the EVT-based adaptive quantile loss function predicts $7.26\%$, which closely matches the empirical tail portion of $7.49\%$.
%Notably, as the number of samples, including extreme values, increases, the PDF of the predicted values from the GRU with EVT-based adaptive quantile loss aligns more closely with the PDF of the empirical values. This improvement is attributed to the relatively low number of tail samples in the overall dataset, underscoring the dependence of the model on effective training to accurately generate these tail predictions.

\begin{figure}[h]
\centering
     \begin{center}

            \label{fig:mae_tail}
            \includegraphics[width=7cm,height=4cm]{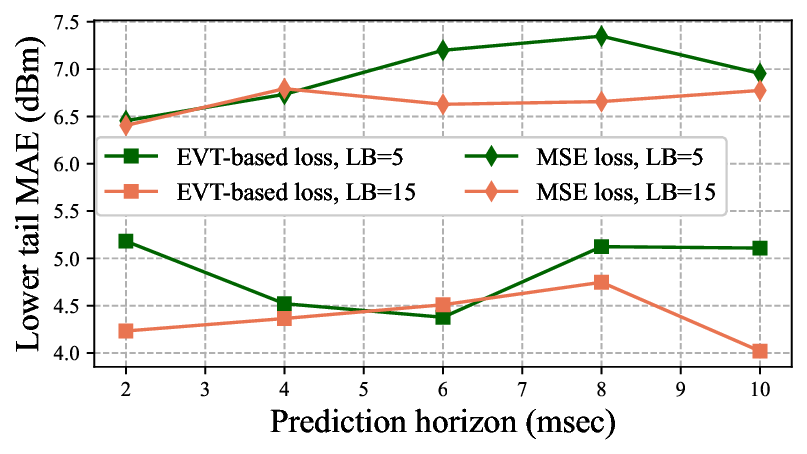}
    \end{center}
    %\vspace{-5mm}
    \caption{MAE results for GRU using MSE-based and EVT-based adaptive quantile losses for tail values across different prediction horizons.}
   \label{fig:mae}
\end{figure}

Fig.~\ref{fig:mae} illustrates the MAE for the tail values across different prediction horizons. As shown in this figure, the MAE for the GRU model with the EVT-based adaptive quantile loss function is approximately $3$~dBm lower than that for the GRU with the MSE-based loss function. This indicates that the primary challenge of predicting tail values within a few milliseconds has been effectively addressed.
Additionally, for a LB period of $15$, the loss is slightly lower than that for a LB of $5$. However, this negligible difference suggests the outperformance of the GRU with the proposed EVT-based adaptive quantile loss function to estimate tail values using the samples collected only within the last $10$~msec. On the other hand, our results show that the GRU model with the MSE loss function outperforms the GRU model with the EVT-based adaptive quantile loss function by less than $0.5$~dBm in terms of MAE for predicting average statistic channel, i.e., received power values above the threshold of $-49$~dBm. 
%\vspace{-0.5cm}
\section{Conclusions}
\label{sec:conclusions}
In this letter, we propose a DRNN with GRU model using an EVT-based adaptive quantile loss function for the channel prediction in URLLC. By incorporating EVT to estimate the GPD parameters fitted to the channel tail distribution, we formulate an adaptive quantile loss function for the DRNN with the GRU model, enhancing its ability to predict extreme values. 
%This approach employs a GAN-EVT model to estimate the shape and scale parameters of the GPD, ensuring a robust reconstruction of the distribution of tail events. By integrating these parameters into an EVT-based adaptive quantile loss function, our model effectively emphasizes tail predictions within a few milliseconds while maintaining accuracy for near-average values. 
Our approach demonstrates improved predictive performance for extreme values as validated through quantitative metrics of MAE and probability density, and visual comparisons with traditional GRU model using the MSE-based loss function.
In the future, to further improve the channel prediction performance of the proposed scheme, we plan to increase the number of training samples using GAN-based data augmentation, and optimize the weights assigned to the loss
of near-average samples and the tail samples. We also plan to account for non-stationarity in the dataset by estimating GPD parameters in real-time and integrating them into the EVT-based adaptive quantile loss function. Moreover, we plan to utilize the channel prediction results for the link adaptation to achieve URLLC.

%\vspace{-5mm}
\balance 

\ifCLASSOPTIONcaptionsoff
  \newpage
\fi
%\vspace{-3mm}
\bibliographystyle{ieeetr}
\bibliography{ChannelPrediction.bib}

\end{document}